\documentclass[aps,prd,twocolumn,superscriptaddress,longbibliography,nofootinbib]{revtex4-2}
\usepackage[utf8]{inputenc}
\usepackage{color}
\usepackage{graphicx}
\usepackage{subfigure}
\usepackage[table]{xcolor}
\usepackage[normalem]{ulem}
\usepackage[pdftex,breaklinks,colorlinks,linkcolor=blue,citecolor=teal,anchorcolor=red,urlcolor=cyan]{hyperref}
\usepackage{orcidlink}
\usepackage{multirow}
\usepackage{subfigure}
\usepackage{gensymb}
\usepackage{amsmath,amssymb,amsfonts}
\usepackage{booktabs} 

\def\prl{Phys. Rev. Lett.}
\def\prd{Phys. Rev. D}

\def\apj{Astrophys. J.}
\def\apjs{Astrophys. J. Suppl.}
\def\apjl{Astrophys. J. Lett.}

\def\pr{Phys. Rev.}

\def\pau_p{Prog. Theor. Phys.}

\def\mnras{Mon. Not. R. Astron. Soc.}

\def\physrep{Phys. Rep.}

\def\sovast{Soviet Ast.}
\def\jcap{J. Cosmology Astropart. Phys}
\def\araa{Annu. Rev. Astron. Astrophys.}

\def\NGW{{\mathcal N}_{\rm GW}}

\begin{document}

\title{Constraining the nuclear equation of state from orbits of primordial black holes inside neutron stars}

\author{Anna Chen}
\email{achen4@bowdoin.edu}
\affiliation{Department of Physics and Astronomy, Bowdoin College, Brunswick, Maine 04011, USA}

\author{Juan Diego DelPrado\orcidlink{0009-0004-2733-733X}}
\email{jdelprado@bowdoin.edu}
\affiliation{Department of Physics and Astronomy, Bowdoin College, Brunswick, Maine 04011, USA}

\author{Thomas W.~Baumgarte\orcidlink{0000-0002-6316-602X}}
\email{tbaumgar@bowdoin.edu}
\affiliation{Department of Physics and Astronomy, Bowdoin College, Brunswick, Maine 04011, USA}

\author{Stuart L.~Shapiro\orcidlink{0000-0002-3263-7386}}
\email{slshapir@illinois.edu}
\affiliation{Department of Physics, University of Illinois at Urbana-Champaign, Urbana, Illinois 61801}
\affiliation{Department of Astronomy and NCSA, University of Illinois at Urbana-Champaign, Urbana, Illinois 61801}

\begin{abstract}
Lacking terrestrial experimental data, our best constraints on the behavior of matter at high densities up to and above nuclear density arise from observations of neutron stars.  Current constraints include those based on measurements of stellar masses, radii, and tidal deformabilities.   Here we explore how orbits of primordial black holes -- should they exist -- inside neutron stars could provide complementary constraints on the nuclear equation of state (EOS).  Specifically, we consider a sample of candidate EOSs, construct neutron star models for these EOSs, and compute orbits of primordial black holes inside these stars.  We discuss how the pericenter advance of eccentric orbits, i.e.~orbital precession, results in beat phenomena in the emitted gravitational wave signal.  Observing this beat frequency could constrain the nuclear EOS and break possible degeneracies arising from other constraints, as well as provide information about the host star.
\end{abstract}


\maketitle

\section{Introduction}
\label{sec:Intro}

The behavior of matter at densities higher than nuclear density remains poorly understood.  As we briefly review in Section \ref{sec:EOS} below, different theoretical approaches make predictions by adopting many-particle quantum field theory prescriptions with different assumptions and approximations, resulting in different candidate nuclear equations of state (EOSs).  Since these high densities cannot be produced in terrestrial laboratories (at least not on macroscopic scales), it is difficult to test these predictions experimentally. Instead, our strongest constraints on the cold nuclear EOS arise from astronomical observations of neutron stars (NSs; see, e.g., \cite{OezF16,Lat21} for reviews).

Current constraints are based primarily on three different types of observations, namely measurements of the stellar mass, the stellar radius, and the tidal deformability (see Section \ref{sec:constraints} below).  In fact, the recent determination of the high mass of the pulsar J0740+6620 ($M = 2.08 M_{\odot}$) using radio observations of the Shapiro time delay \cite{Croetal20,Fonetal21}, together with measurements of NS radii using the Neutron Star Interior Composition Explorer (NICER, e.g.~\cite{Miletal19,Raaetal21,Ditetal24}) and constraints on the tidal deformability from the gravitational-wave (GW) signal GW170817 observed by the Laser Interferometer Gravitational Wave Observatory (LIGO, e.g.~\cite{Abbetal19,Abbetal20}) have already ruled out several candidate EOSs.  The latter constraints in particular are likely to improve even in the near future if more binary NS mergers are observed.  

In this paper we explore a hypothetical and speculative scenario that might, however, provide independent and complementary constraints on the nuclear EOS.  Specifically, we consider orbits of primordial black holes (PBHs) inside NSs, and demonstrate how the characteristics of GW signals emitted by these PBHs could help determine both the EOS and the properties of the host star.

First proposed by Zel'dovich and Novikov \cite{ZelN67} and Hawking \cite{Haw71} (see also \cite{CarH74}), PBHs may have formed in the early Universe, and may either contribute to, or even make up, the dark-matter content of the Universe.  While observational constraints limit the contribution of PBHs to the dark matter (DM) in some mass ranges, they remain plausible in others (see, e.g., \cite{Khl10,CarKSY21,CarK20,CarCGHK24} for reviews, see also \cite{MonCFVSH19}).  Most relevantly for our purposes here, PBHs with masses $10^{-16} M_\odot \lesssim m_{\rm PBH} \lesssim 10^{-10} M_\odot$ are completely unconstrained and could make up the entire DM content, while PBHs with masses $M \simeq 10^{-6} M_\odot$ could contribute up to about 10\% of the DM (see, e.g., Fig.~10 in \cite{CarKSY21}).  

If PBHs exist, then some are likely to interact with other celestial objects.   A number of authors have invoked such interactions, especially direct collisions, as explanations for or sources of a variety of different astronomical phenomena (see, e.g., \cite{BauS24b} for examples and references).  If the PBH loses a sufficient amount of energy in the interaction with a star, it may end up being bound gravitationally.  Once captured, the PBH is likely to lose more energy and ultimately be swallowed entirely, resulting in the PBH acting as an intruder in the host star.  Due to their high densities, NSs turn out to be the most efficient at extracting energy from a PBH orbit, and are therefore the most likely to act like such a host and have a PBH embedded inside (see, e.g., \cite{CapPT13,AbrBW18,GenST20,AbrBUW22,BauS24b,CaiBK24} and references therein).  Inside the NS, an ``endoparasitic" PBH (borrowing the language of \cite{EasL19}) slowly spirals to the center of the NS, accretes stellar material, and ultimately induces dynamical collapse of the host star (see \cite{EasL19,RicBS21b,SchBS21,BauS24c} for numerical simulations). 

During its inspiral toward the stellar center, the PBH completes numerous orbits about the center and emits continuous GWs in the process.  Several authors have previously suggested that the characteristics of these GWs might reveal information about the structure of the star and hence the nuclear EOS \cite{HorR19,ZouH22,GaoDGZZZ23}.  While the discussion in \cite{ZouH22,GaoDGZZZ23} focused on circular orbits, circular orbits appear to be unlikely (see \cite{BauS24b} as well as \cite{Hof85,SzoML22}).  However, the precession of noncircular orbits and associated GW beats can reveal significantly more information about the stellar structure than circular orbits alone (see \cite{BauS24} and Section \ref{sec:review} below).  

In \cite{BauS24}, two of us explored these effects assuming a simple polytropic NS EOS and found that the rate of pericenter advance, and hence the GW beat frequency, strongly depends on the polytropic index and therefore on the stiffness of the EOS.   In this paper we instead consider a number of different candidate nuclear EOSs and examine how a hypothetical future observation of GWs from a PBH orbiting inside a NS could be used to determine both the EOS and properties of the host star, including its mass.  

Event rates for collisions between PBHs and NSs are estimated to be small (see, e.g., \cite{CapPT13,MonCFVSH19,Abretal09,HorR19,ZouH22}, as well as Section I of the Supplemental Material in \cite{BauS24}), so that the detection of such a GW signal will most likely be possible only with more sensitive GW detectors than those currently available.  However, these estimates depend on a number of assumptions, and the event rates could be more favorable in special environments (e.g.~galactic centers or AGNs).  Alternative scenarios could also lead to small black holes inside NSs: PBHs could be captured by processes other than collisions (e.g.~\cite{BamSDFV09,GenST20,HorR19,PanL14}), or other DM particles could accumulate inside NSs and collapse to form small black holes there (e.g.~\cite{GolN89,McDYZ12,BraLT18,EasL19}).  

Our paper is organized as follows.  In Section \ref{sec:NSmodels} we discuss our sample of candidates of realistic nuclear EOSs, their implementation as piecewise polytropes, models of NSs based on these EOSs, and the calculation of PBH orbits inside such NSs.  In Section \ref{sec:constraints} we review existing constraints on nuclear EOSs based on their maximum allowed mass, stellar radius, and tidal deformability.   In Section \ref{sec:PBH} we then demonstrate how orbits of PBHs inside NSs may provide new constraints.  Specifically, we review the effects of orbital precession and GW beats in Section \ref{sec:review}, and explore how future observations of these GW beats could provide sensitive and complementary constraints on the nuclear EOS in Section \ref{sec:scenarios}.  We briefly summarize our conclusions in Section \ref{sec:discussion}.

\section{Models of NSs}
\label{sec:NSmodels}

\subsection{Nuclear EOSs}
\label{sec:EOS}

\begin{figure}
    \centering
    \includegraphics[width=0.95\linewidth]{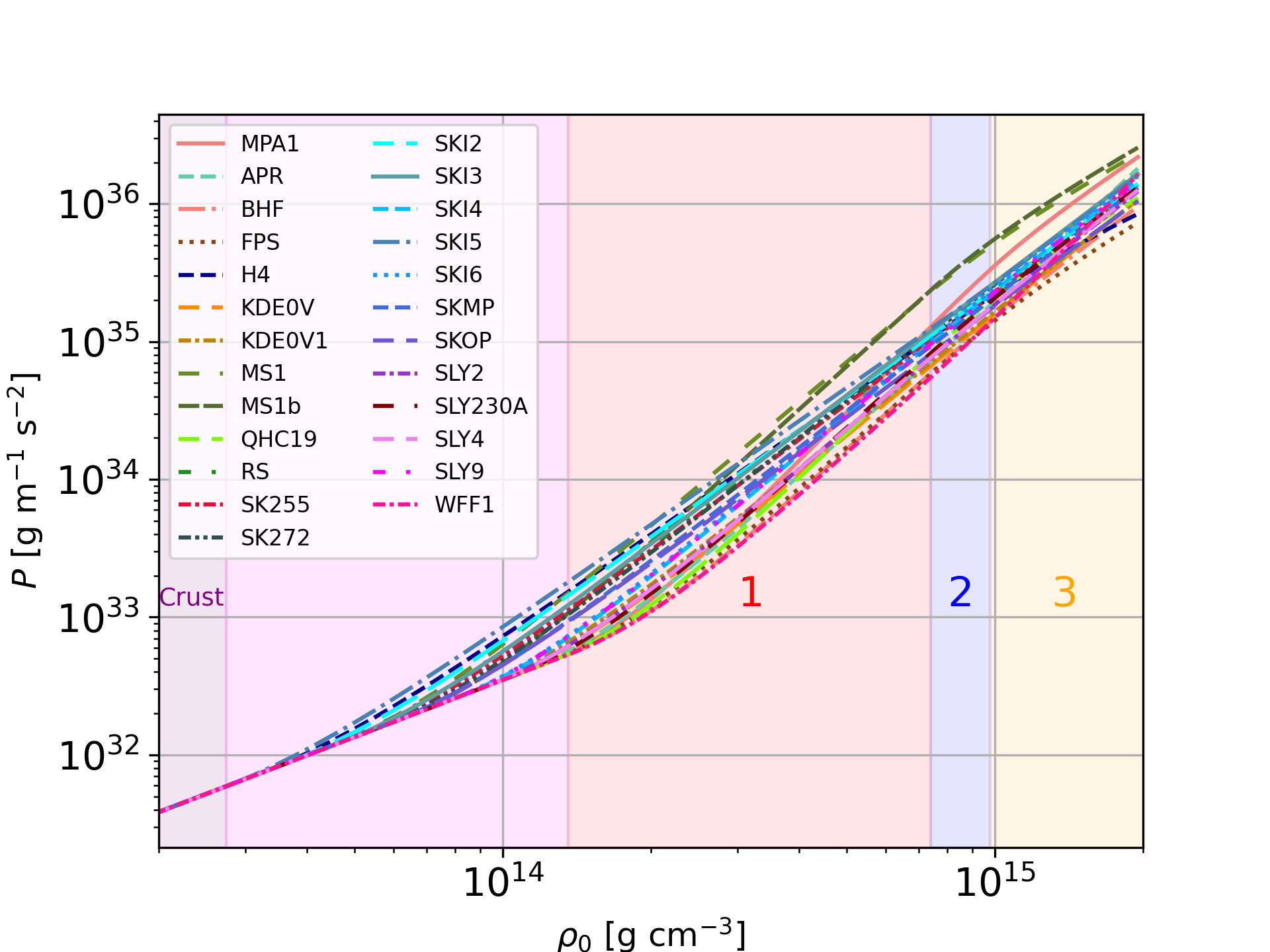}
    \caption{Graphs of the pressure $P$ versus rest-mass density $\rho_0$ for our sample of EOSs, shown primarily to demonstrate the range of predictions.  The colors in the background indicate the different regions in our piecewise polytrope approximation; the unlabeled pink region shows the range of the boundary density $\rho_{01}$ between the crust and the region 1.}
    \label{fig:PvrsRho}
\end{figure}

We consider the same sample of candidate EOSs as in \cite{OBoyMSR20}, which, in turn, is based on the catalogues assembled in \cite{Abbetal20} as well as the CompOSE \cite{CompOSE} and Oezel EOS projects \cite{Oezel}.  We list the different EOSs in Tables \ref{tab:EOS_parameters} and \ref{tab:EOS_summary}, and graph the corresponding relations between the pressure $P$ and the rest-mass density $\rho_0$ in Fig.~\ref{fig:PvrsRho}.  The different EOSs are constructed using different approximations for implementing many-particle quantum field theory, e.g.~relativistic or non-relativistic mean-field theory, liquid-drop models, Hartree-Fock methods, Brueckner-Bethe-Goldstone methods, or Wigner-Seitz cells.  Some models adopt a parametrized description of the force between the EOS constituents, and different versions of the EOSs use different values of the nuclear symmetry energy and incompressibility.  Most of the EOSs in our sample are based on pure hadronic matter, while some also allow for more exotic particles like hyperons or kaons. We refer to these candidate EOSs with the abbreviations adopted in \cite{OBoyMSR20}. 

We include several EOSs for plain $npe\mu$ nuclear matter:
\begin{itemize}
    \item three variational-method EOSs (APR, FPS, and WFF1, see \cite{ReaLOF09, CompOSE}),
    
    \item one non-relativistic (BHF) and one relativistic (MPA1) Brueckner-Hartree-Fock EOS, the former without muons (see \cite{ReaLOF09, Abbetal20, CompOSE}),
    
    \item fifteen EOSs using Brueckner-Hartree-Fock methods, Skyrme forces, and Wigner-Seitz cells (KDE0V, KDE0V1, RS, SK255, SK272, SKI2-6, SKMP, SKOP, SLY2, SLY230A, and SLY9, see \cite{Abbetal20, CompOSE}),
    
    \item and two relativistic mean-field theory EOSs (MS1, MS1b, see \cite{ReaLOF09, Abbetal20}).
\end{itemize}
    
We also include several EOSs including exotic particles:
\begin{itemize}
    \item one relativistic mean field theory EOS with hyperons (H4, see \cite{ReaLOF09, Abbetal20}),

    \item one mixed EOS using variational-method nuclear matter and Nambu–Jona–Lasinio modeled quark matter, with hyperons but without muons (QHC19, see \cite{CompOSE, Abbetal20}),
    
    \item and one EOS using Brueckner-Hartree-Fock methods, Skyrme forces, and Wigner-Seitz cells, which includes hyperons (SLY4, see\cite{CompOSE}).
\end{itemize}
We refer to \cite{Abbetal20} (in particular their Table A1) as well as \cite{CompOSE,Oezel,ReaLOF09} for more details and references.  

We note that some pairs of EOSs, most notably SKI4 and SLY9 as well as SLY2 and SLY4, result in very similar relations between $P$ and $\rho_0$ and are difficult to distinguish in a graph like Fig.~\ref{fig:PvrsRho} (see also the corresponding parameters in Table~\ref{tab:EOS_parameters}).  Not surprisingly, these EOSs will also be difficult to distinguish with any of the constraints that we discuss below.

\subsection{Piecewise polytropes}
\label{sec:PWP}

\begin{table}[]
    \centering
    \begin{tabular}{c|c|c|c|c|c}
       EOS  &  $\log \rho_{01}$ & log $K_{1}$ & $\Gamma_{1}$ & $\Gamma_{2}$ & $\Gamma_{3}$    
       \\
       \hline
       APR 
         &-3.7517  & 5.3780 & 3.169 & 3.452 & 3.310 
        \\
       BHF &-3.6607  & 5.6180 & 3.284 & 2.774 &   2.616
       \\
       FPS & -3.7037 & 5.2116 & 3.147 & 2.652 &   2.120
       \\
       H4 & -4.2917 & 3.6251 & 2.514 & 2.333 & 1.562 
       \\
       KDE0V & -3.8127 & 4.7443 & 2.967 & 2.835 & 2.803
       \\
       KDE0V1 & -3.8617 & 4.5703 & 2.900 & 2.809 & 2.747 
       \\
       MPA1 & -3.7027 & 7.0578 & 3.662 & 3.057 & 2.298 
       \\
       MS1 & -4.1337 & 5.3758 & 2.998 & 2.123 & 1.955 
       \\
       MS1b & -3.9957 & 6.0950 & 3.241 & 2.136 & 1.963 
       \\
       QHC19 & -3.6887 & 6.1567 & 3.419 & 2.760 & 2.017  
       \\
       RS & -4.1497 & 3.9556 & 2.636 & 2.677 & 2.647 
       \\
       SK255 & -4.1117 & 4.1297 &  2.693 & 2.729 &  2.667 
       \\
       SK272 & -4.0587 & 4.4974 & 2.804 & 2.793 & 2.733 
       \\
       SKI2 & -4.2387 & 3.8183 & 2.575 & 2.639 & 2.656 
       \\
       SKI3 & -4.1307 & 4.3031 & 2.729 & 2.680 & 2.708 
       \\
       SKI4 & -3.8837 & 5.0893 & 3.029 & 2.759 & 2.651 
       \\
       SKI5 & -4.3527 & 3.6660 & 2.505 & 2.708 & 2.727 
       \\
       SKI6 & -3.8887 & 5.1329 & 3.036 & 2.762 & 2.653 
       \\
       SKMP & -4.0277 & 4.3024 & 2.766 & 2.741 & 2.698 
       \\
       SKOP & -4.0296 & 4.0306 & 2.693 & 2.660 &  2.579
       \\
       SLY2 & -3.8237 & 4.9740 & 3.026 & 2.871 & 2.760 
       \\
       SLY230A & -3.7696 & 5.4699 & 3.184 & 2.895 & 2.588 
       \\
       SLY4 & -3.8107 & 5.0320 & 3.045 & 2.884 & 2.773 
       \\
       SLY9 & -3.8917 & 5.0133 & 3.005 & 2.796 & 2.652 
       \\
       WFF1 & -3.6577 & 5.4572 & 3.240 & 3.484 & 3.695
    \end{tabular}
    \caption{Values of the generalized PP parameters for our catalogue of EOSs (see Table III in \cite{OBoyMSR20}).  The exponents $\Gamma_i$ are dimensionless, and values for $\rho_{01}$ and $K_i$ are provided in geometrized units of solar mass, i.e.~$[\rho_{0}] = M_{\odot}^{-2}$, and $[K_{1}] = M_{\odot}^{2\Gamma_{1}-2}$.  All other parameters can be computed from those listed here using the continuity conditions discussed in the text.}
    \label{tab:EOS_parameters}
\end{table}

We approximate the nuclear EOSs of Section \ref{sec:EOS} using the {\em generalized piecewise polytrope} approach of \cite{OBoyMSR20}, which is based on the earlier piecewise polytrope approximation of \cite{ReaLOF09}.  Specifically, we write the pressure $P$ as a function of the rest-mass density $\rho_0$ as
\begin{equation} \label{P_PP}
P = K_i \rho_0^{\Gamma_i} + \lambda_i
\end{equation}
and the specific internal energy density $\epsilon$ as 
\begin{equation} \label{epsilon_PP}
\epsilon = \frac{K_i}{\Gamma_i - 1} \rho_0^{\Gamma_i - 1} + b_i + \frac{\lambda_i}{\rho_0},
\end{equation}
where the constants $K_i$, $\Gamma_i$, $\lambda_i$, and $b_i$ are chosen in three different high-density regions in such a way that $P$, $\epsilon$, as well as the derivative $dP /d\rho_0$ are continuous across the boundaries of these regions (see \cite{OBoyMSR20} for details).\footnote{The generalized piecewise polytrope approach of \cite{OBoyMSR20}, which we adopt here, differs from the original version of \cite{ReaLOF09} in that the latter did not include the terms $\lambda_i$ in (\ref{P_PP}) and (\ref{epsilon_PP}).  Without these terms, the pressure $P$ can be made continuous across the boundaries of regions, but not its derivative, resulting in discontinuities in the speed of sound.}  A further fitting parameter is the rest-mass density $\rho_{01}$ at the interface between the crust and region 1.  The rest-mass densities at the two other interfaces are the same for all EOSs and are chosen as in \cite{OBoyMSR20}, namely $\rho_{02} = 10^{14.87} \mbox{g}\,\mbox{cm}^{-3}$ and $\rho_{03} = 10^{14.99} \mbox{g}\,\mbox{cm}^{-3}$.   In our approach chosen here we join the high-density part of the EOS to the SLY4 EOS in the low-density crust (see Table II of \cite{OBoyMSR20}), which results in the parameters provided in Table III of \cite{OBoyMSR20} (which we reproduce in Table \ref{tab:EOS_parameters}).  In Fig.~\ref{fig:PvrsRho} we have included a color-coding that indicates the different density regions adopted in the piecewise polytrope approximation.

\subsection{Stellar models}
\label{sec:OV}

\begin{figure*}[t]
        \centering
        \includegraphics[width=0.45\textwidth]{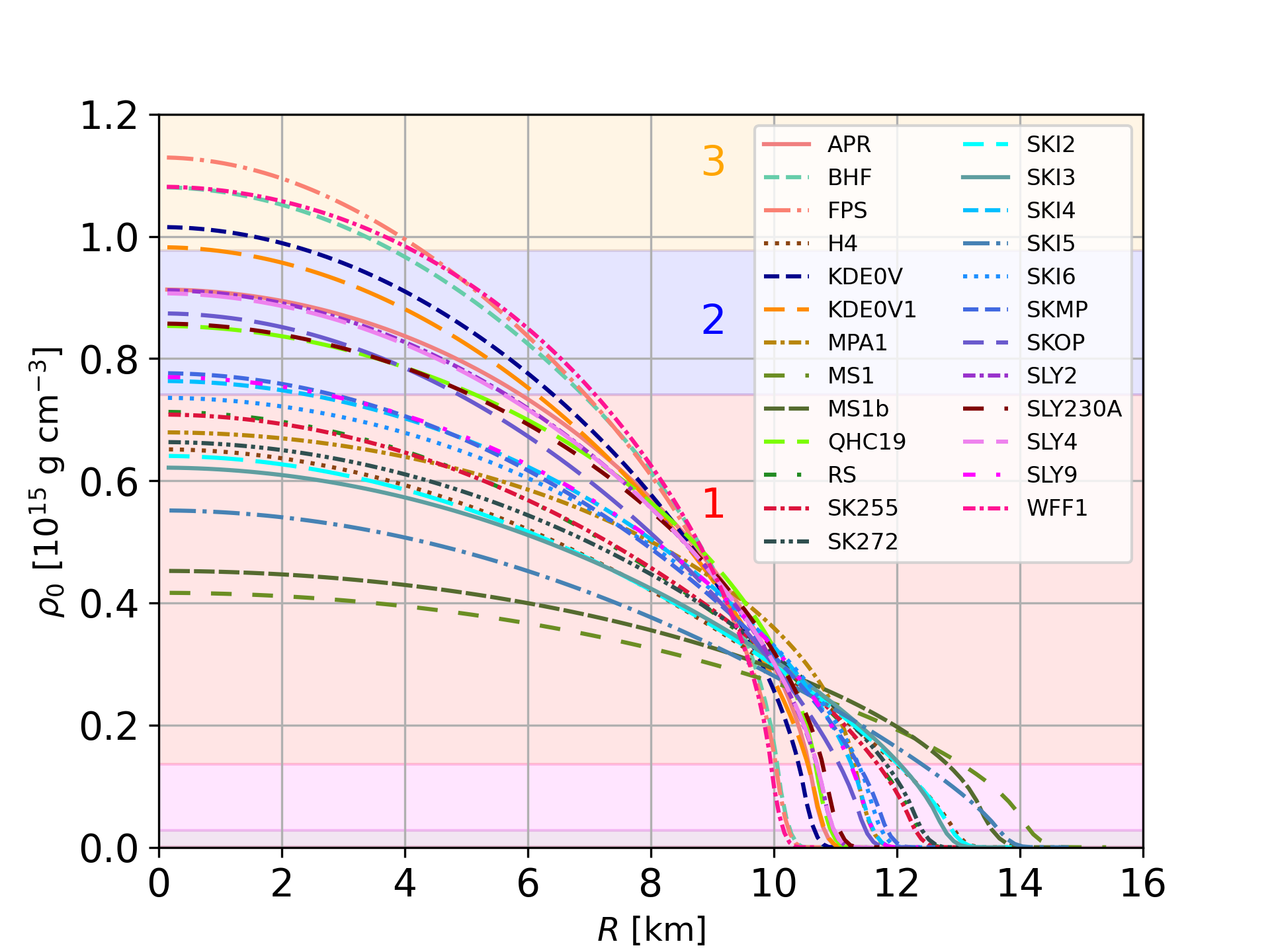}
        \includegraphics[width=0.45\textwidth]{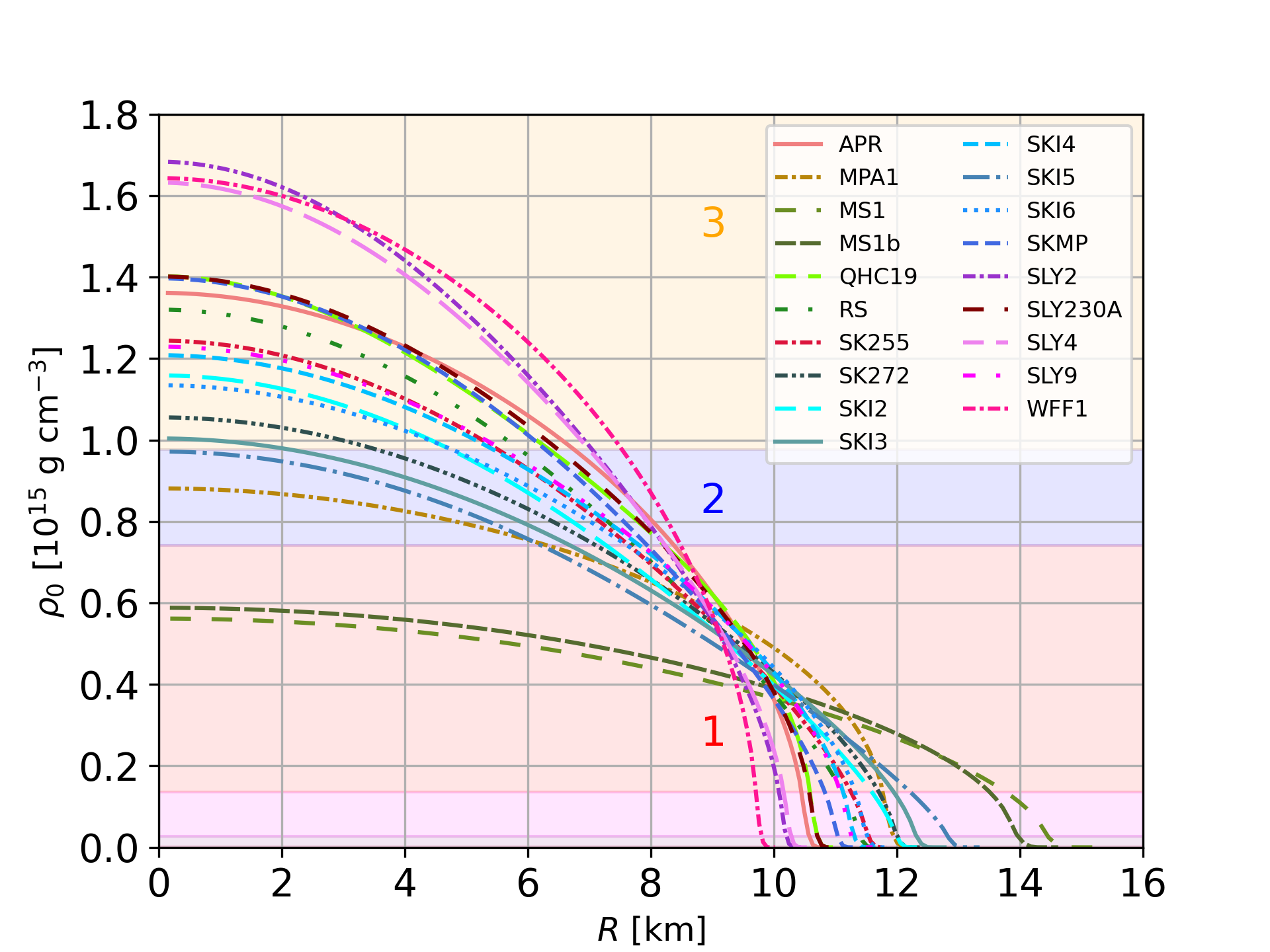}
        \caption{Density profiles for a NS with gravitational mass $M = 1.4 M_\odot$ (left) and $M = 2.0 M_\odot$ (right).  We graph the rest-mass density $\rho_0$ as a function of areal radius $R$. for the EOSs listed in Tables~\ref{tab:EOS_parameters} and \ref{tab:EOS_summary}.  The background color shading indicates the different density regions adopted in the generalized piecewise polytrope approximation; the pink region without a label shows the range of the (fitted) boundary density $\rho_{01}$ at the interface between the crust and region 1.}
        \label{fig:GPP_rho_0vsR_1.4}
\end{figure*}

\begin{figure*}
        \centering
        \includegraphics[width=0.48\textwidth]{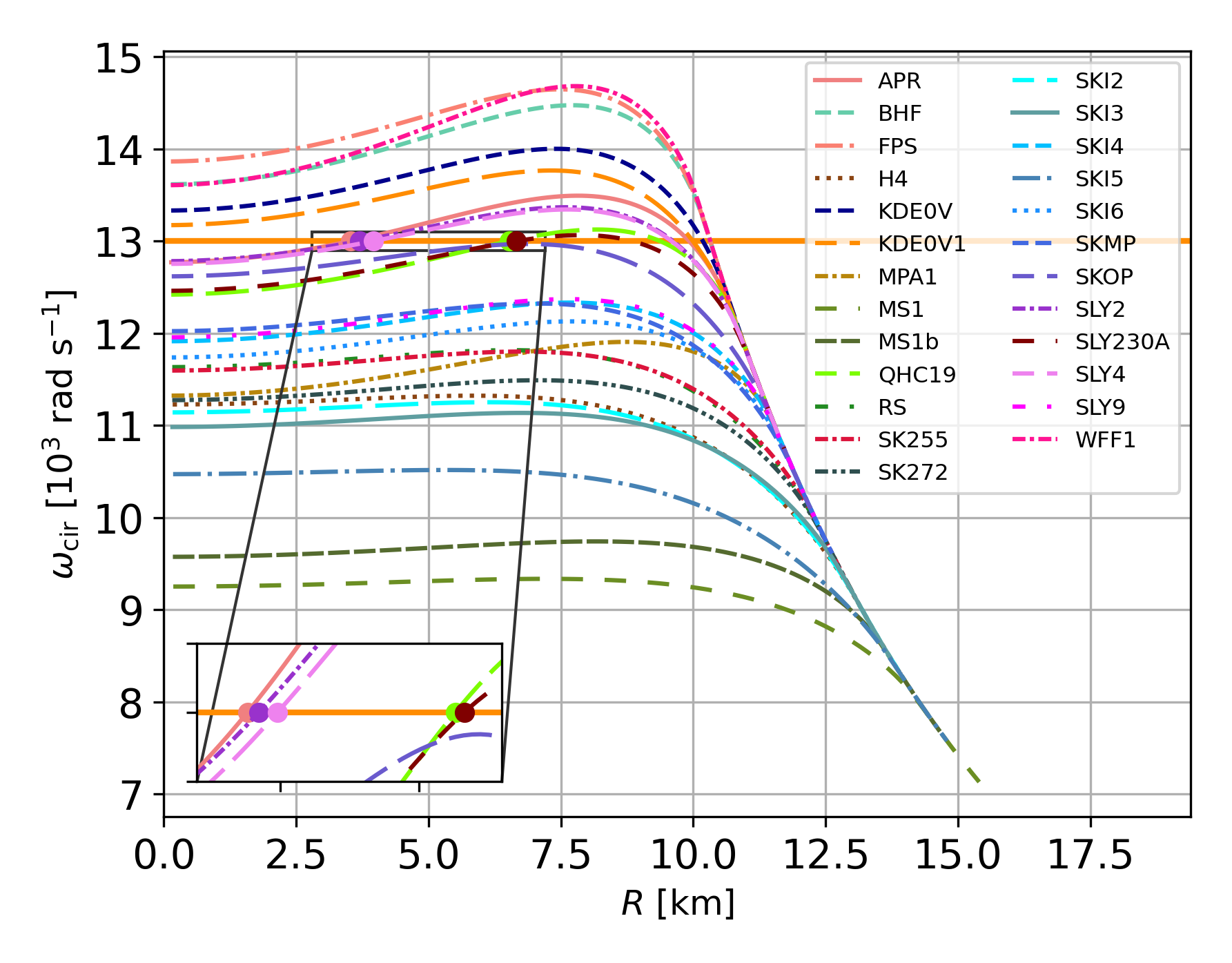}
        \includegraphics[width=0.48\textwidth]{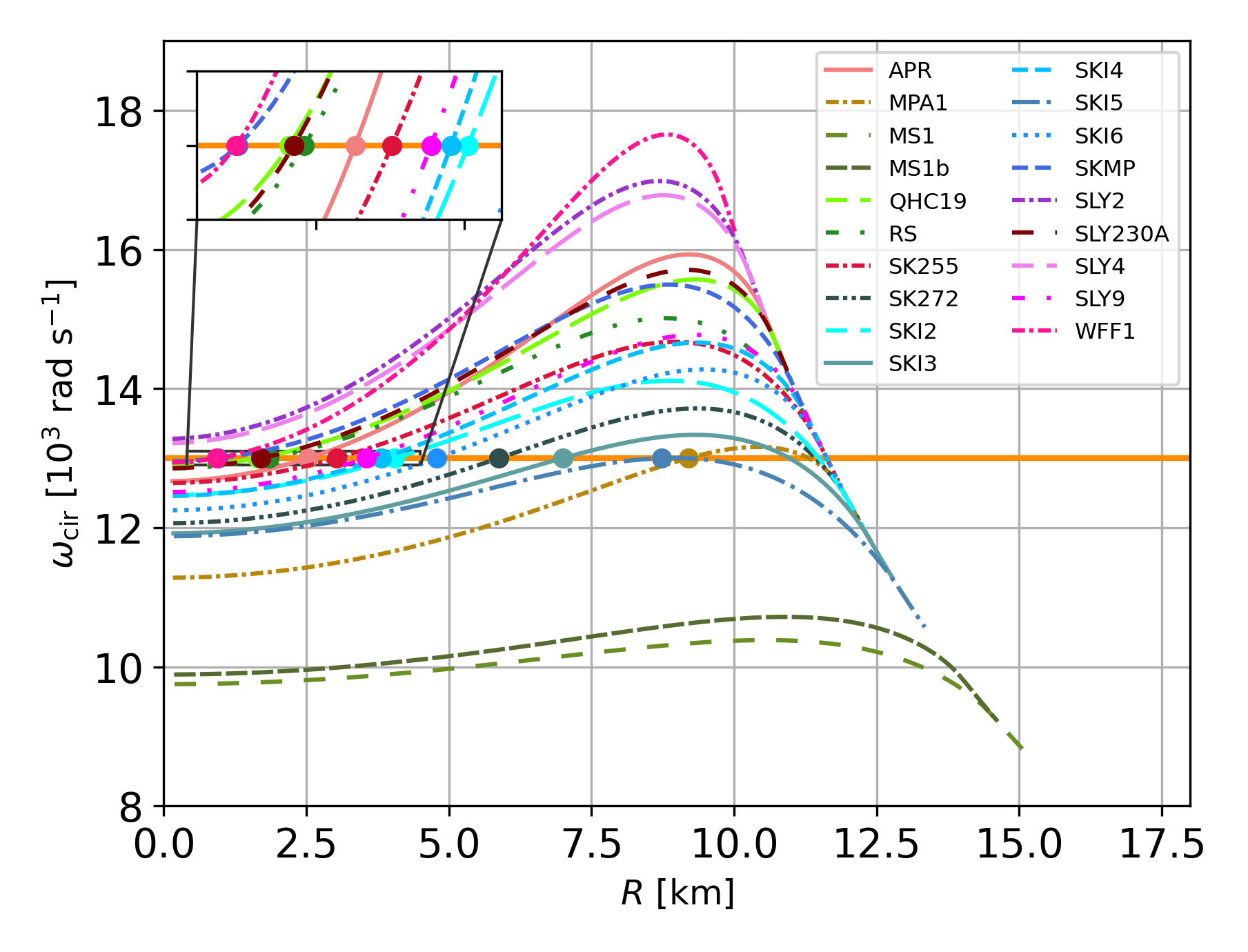}
        \caption{Angular speed $\omega_{\rm cir}$ of a primordial black hole in circular orbit inside NSs of mass $M = 1.4 M_\odot$ (left panel) and $M = 2.0 M_\odot$ (right panel). We graph $\omega_{\rm cir}$ as a function of the orbital (areal) radius $R$ for the same set of EOSs as in Fig.~\ref{fig:GPP_rho_0vsR_1.4}.  The horizontal (orange) line marks an angular speed $\omega = 1.3 \times 10^4\,\mbox{rad}\,\mbox{s}^{-1}$ identified in a hypothetical GW observation, as discussed in Section \ref{sec:PBH}. }
        \label{fig:GPP_omegavsR_1.4}
\end{figure*}


\begin{table*}[]
    \centering
    \begin{tabular}{c|c|c|c|c|c|c|c|c|c|c|c}
       & & \multicolumn{4}{|c|}{$M = 1.4 M_\odot$} & \multicolumn{4}{|c}{$M = 2.0 M_\odot$} \\
       \hline
       EOS  &  $M_{\rm max}~[\rm{M}_{\odot}]$   & $R_{*}$ [km] & $\Lambda$ & $\mathcal{N}_{\rm{GW}}$ & $\omega_{\rm max}$ [rad s$^{-1}$]  & $R_{*}$ [km] & $\Lambda$ & $\mathcal{N}_{\rm{GW}}$ & $\omega_{\rm max}$ [rad s$^{-1}$]
       \\
       \hline \hline 
       APR 
         & 2.167 & \cellcolor[HTML]{FFFFFF}11.43 & 273.3 &17.38   & 13.49 &  10.85 & 14.67 & 18.72 & 15.93
        \\
       \hline 
       BHF & \cellcolor[HTML]{FFFFFF}1.904 & \cellcolor[HTML]{FFFFFF}10.79 & 190.2 & -& 14.47 &    - & - &- & -
       \\
       \hline
       FPS & \cellcolor[HTML]{FFFFFF}1.778 & \cellcolor[HTML]{FFFFFF}10.77 & 180.9 & - & 14.65 &   - & - &- & -
       \\
       \hline
       H4 & \cellcolor[HTML]{fFFFFF} 1.998 & \cellcolor[HTML]{FFFFFF}13.92 & \cellcolor[HTML]{FFFFFF}749.9 & -& 11.32 &   - & - &- & -
       \\
       \hline
       KDE0V & \cellcolor[HTML]{FFFFFF}1.948 & \cellcolor[HTML]{FFFFFF}11.25 & 231.9 & -  & 14.00 &  - & - &- & -
       \\
       \hline
       KDE0V1 & \cellcolor[HTML]{fFFFFF}1.952 & \cellcolor[HTML]{FFFFFF}11.45 & 255.1 & - & 13.77 &   - & - &- & -
       \\
       \hline 
       MPA1 & 2.448 & \cellcolor[HTML]{fFFFFF}12.35 & \cellcolor[HTML]{fFFFFF}486.1 & -  & 11.91 &  12.36 & 46.75 & 1.884 & 13.16
       \\
       \hline
       MS1 & 2.704 & \cellcolor[HTML]{FFFFFF}15.40 & \cellcolor[HTML]{FFFFFF}1623 & -  & 9.335 &  15.18 & 171.1 &- & 10.38
       \\
       \hline
       MS1b & 2.746 & \cellcolor[HTML]{FFFFFF}14.66 & \cellcolor[HTML]{FFFFFF}1281 & - & 9.741 &   14.62 & 142.4 &- & 10.72
       \\
       \hline 
       QHC19 & 2.068 & \cellcolor[HTML]{fFFFFF}11.56 & 307.7 & 4.995  & 13.12 & 11.00 & 16.60 & 42.93 & 15.57
       \\
       \hline
       RS & 2.081 & \cellcolor[HTML]{FFFFFF}13.23 & \cellcolor[HTML]{fFFFFF}581.6 & -  & 11.82 &  11.77 & 21.48 & 36.24  & 15.01
       \\
       \hline
       SK255 & 2.112 & \cellcolor[HTML]{FFFFFF}13.16 &  \cellcolor[HTML]{fFFFFF}575.4 & -   & 11.80 &  11.91 & 24.71 & 14.32 & 14.67
       \\
       \hline 
       SK272 & 2.210 & \cellcolor[HTML]{FFFFFF}13.29 & \cellcolor[HTML]{fFFFFF}637.1 & - & 11.49 &   12.43 & 37.04 & 4.311 & 13.71
       \\
       \hline 
       SKI2 & 2.125 & \cellcolor[HTML]{FFFFFF}13.87 & \cellcolor[HTML]{FFFFFF}755.1 & -  & 11.25 &  12.42 & 31.99 & 8.478 & 14.12
       \\
       \hline
       SKI3 & 2.219 & \cellcolor[HTML]{FFFFFF}13.73 & \cellcolor[HTML]{FFFFFF}755.7 & -  & 11.14 &  12.77 & 44.12 & 3.080 & 13.34
       \\
       \hline 
       SKI4 & 2.152 & \cellcolor[HTML]{fFFFFF}12.36 & \cellcolor[HTML]{fFFFFF}433.1 & - & 12.33 &   11.66 & 24.37 & 9.243  & 14.66
       \\
       \hline
       SKI5 & 2.199 & \cellcolor[HTML]{FFFFFF}14.79 & \cellcolor[HTML]{FFFFFF}1079 & -  & 10.52 &  13.34 & 53.38 & 1.828  & 13.00
       \\
       \hline
       SKI6 & 2.188 & \cellcolor[HTML]{fFFFFF}12.51 & \cellcolor[HTML]{fFFFFF}469.3 & - & 12.13 &   11.89 & 28.72 & 6.214  & 14.28
       \\
       \hline
       SKMP & 2.075 & \cellcolor[HTML]{fFFFFF}12.62 & \cellcolor[HTML]{fFFFFF}456.3 & - &  12.32 & 11.36 & 17.31 & 127.3  & 15.50
       \\
       \hline
       SKOP & \cellcolor[HTML]{fFFFFF} 1.955 & \cellcolor[HTML]{fFFFFF}12.21 & \cellcolor[HTML]{fFFFFF}359.8 & -  & 12.97 &  - & - & - & -
       \\
       \hline
       SLY2 & 2.031 & \cellcolor[HTML]{fFFFFF}11.61 & 290.8 & 15.59 & 13.37 &   10.45 & 9.28 & - & 16.99
       \\
       \hline
       SLY230A & 2.091 & \cellcolor[HTML]{fFFFFF}11.72 & 320.34 & 4.694  & 13.06 &  11.00 & 15.73 & 40.02  & 15.71
       \\
       \hline
       SLY4 & 2.041 & \cellcolor[HTML]{fFFFFF}11.62 & 292.8 & 13.55  & 13.34 &   10.54 & 10.11 & - & 16.78
       \\
       \hline
       SLY9 & 2.144 & \cellcolor[HTML]{fFFFFF}12.35 & \cellcolor[HTML]{fFFFFF}428.2 & -  & 12.37 &  11.60 & 23.28 & 10.54  & 14.77
       \\
       \hline
       WFF1 & 2.101 & \cellcolor[HTML]{FFFFFF}10.67 & 177.3 & -  & 14.68 &  10.015 & 7.51 & 108.8 & 17.65
       \\
       \hline
    \end{tabular}
    \caption{Properties of neutron stars constructed from our catalogue of nuclear EOSs.  For each EOS we list the (nonrotating) maximum allowed gravitational mass $M_{\rm  max}$ together with parameters for two different NS models with masses $M = 1.4 M_\odot$ and $M = 2.0 M_\odot$. We list the stellar radius $R_*$ for each one of these models, the (dimensionless) tidal deformability $\Lambda$, the value of $\NGW = \tau_{\rm beat} / \tau_{\rm orb}$ for our fiducial value of $\omega_{\rm orb} = \omega_{\rm fid} = 1.3 \times 10^4$ rad s$^{-1}$ and a nearly circular orbit, as well as the maximum circular orbital frequency $\omega_{\rm max}$.  A blank entry indicates either that the corresponding EOS cannot support a star of mass $M = 2.0 M_\odot$, or that the specific stellar model does not allow circular orbits with angular frequency $\omega_{\rm fid}$.}
    \label{tab:EOS_summary}
\end{table*}


Given an EOS we can construct models of spherically symmetric, nonrotating NSs by solving the Oppenheimer-Volkoff equations of relativistic hydrostatic equilibrium \cite{OppV39}.  Varying the central rest-mass density $\rho_{0c}$ we obtain stars with different gravitational mass $M$. In particular we can identify the (nonrotating) {\em maximum allowed mass} $M_{\rm max}$ for each EOS, which we list in the second column of Table \ref{tab:EOS_summary} (see also Fig.~\ref{fig:M_max} below).

For each EOS we may also construct stellar models for given stellar masses $M$, and hence obtain the (areal) stellar radius $R_*$ of those particular stellar models, as well as profiles of their internal composition.  In Fig.~\ref{fig:GPP_rho_0vsR_1.4} we show density profiles for two specific masses, namely $M = 1.4 M_\odot$ representing a typical NS mass and $M = 2 M_\odot$ representing a high-mass star.  We also list the corresponding stellar radii $R$ in Table \ref{tab:EOS_summary}.

Anticipating our discussion in Section \ref{sec:PBH} below we also compute the orbital frequency $\omega_{\rm cir}$ of a circular orbit of radius $R$ inside the above stellar models, and show the results in Fig.~\ref{fig:GPP_omegavsR_1.4}.
Expressions for the circular orbital frequency $\omega_{\rm cir}$ can be found by setting the radial component of the PBH's four-velocity $u_r$, as well as its time derivative $du_r/dt$, to zero in the relativistic geodesic equations (e.g.~Eqs.~(25) in \cite{BauS24b}) and solving for $\omega_{\rm cir} = d\varphi/dt$.  The result depends on the metric coefficients and their derivatives that were computed for the stellar models of Section \ref{sec:OV}.  Note in particular that, contrary to what one might expect from Newtonian gravity, $\omega_{\rm cir}$ does {\em not} necessarily decrease as $R$ increases.  Instead, most curves first increase starting from a ``central" value $\omega_0$ (for $R \rightarrow 0$) to a maximum value $\omega_{\rm max}$ at finite (non-zero) radius, before decreasing again toward the stellar surface.  The maxima are more pronounced (i.e.~the ratio $\omega_{\rm max}/\omega_0$ is larger) for higher-mass stars, for which the relativistic effects are larger.  We list the values of $\omega_{\rm max}$ for different EOSs and different stellar masses in Table~\ref{tab:EOS_summary}.

\subsection{Orbits of PBHs}
\label{sec:orbits}

For a given NS model we can find orbits of PBHs by solving the relativistic geodesic equations, as we discussed in \cite{BauS24,BauS24b}.   We start these integrations at a time $t = 0$ by placing a PBH at an initial radius $R(0)$ inside the star and setting its initial radial velocity to zero.  We then compute the angular velocity from the orbital frequency $\omega_{\rm cir}$, but reduce this value by a fraction $f_\varphi$ in order to obtain eccentric orbits.  For the nearly-circular orbits considered in Section \ref{sec:PBH} we use $f_\varphi = 0.95$.

The geodesic equations are strictly valid only in the limit that all dissipative forces can be ignored.  In reality, the PBH will lose energy and angular momentum due to dynamical friction, accretion drag, and radiation reaction losses (see, e.g., Section IV in \cite{BauS24b} for a treatment).  For sufficiently small PBHs, $m_{\rm PBH} \ll M$, the secular timescale associated with these dissipative forces is much longer than the dynamical timescale.  Since we integrate the equations for only a few orbits at a time, we may therefore ignore dissipative losses for our purposes here.  In this limit, the orbit of the PBH also becomes independent of the PBH mass $m_{\rm PBH}$.  We do, however, use a Newtonian quadrupole formalism (see \cite{PetM63,Pet64}) to estimate the GW signal emitted by the PBH (see Appendix B in \cite{BauS24b} for details).  The amplitude of this GW signal is proportional to $m_{\rm PBH} R^2$, where $R$ is the radius (or semimajor axis) of the orbit (see Eq.~(59) in \cite{BauS24b}).

\section{Existing constraints on the EOS}
\label{sec:constraints}

In this section we briefly review existing constraints on nuclear EOSs based on the observations of the stellar mass, radii, and tidal deformability.  Other future constraints may also be based on measurements of the stellar moment of inertia (see, e.g.~\cite{MorBSP04,LatS05}).

\subsection{Maximum allowed mass}
\label{sec:m_max}

\begin{figure}
    \centering
    \includegraphics[width=0.95\linewidth]{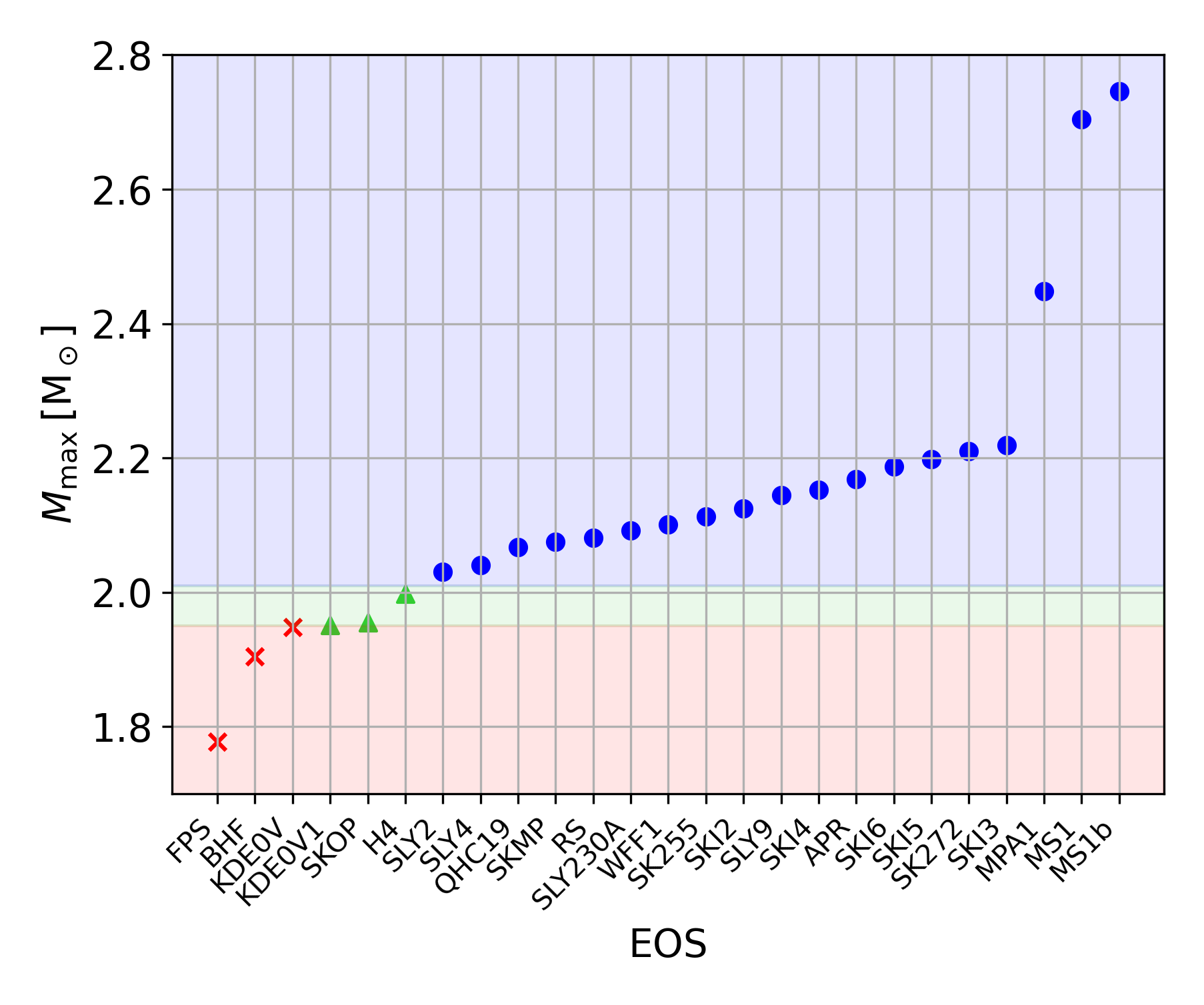}
    \caption{The (nonrotating) maximum allowed gravitational mass $M$ for our sample of EOSs. EOSs marked by (blue) circles satisfy the mass constraint (\ref{mass_constraint}) within the 68\% confidence limits, those marked by (green) triangles are within the 95\% confidence limits, and those marked by (red) crosses are outside those limits. }
    \label{fig:M_max}
\end{figure}

The first constraint on candidate EOSs results from the requirement that it has to be able to support the most massive known NS.  Currently, the largest precisely known NS mass is that of the pulsar J0740+6620,
\begin{equation} \label{M_J0407}
M^{\rm J0740} = 2.08 \pm 0.07 M_\odot 
\end{equation}
(see \cite{Fonetal21}).  Here the error bars correspond to 
$68.3\%$ credibility limits, while the $95.4\%$ limits would allow the mass to be as small as $1.95 M_\odot$.  We now require that the maximum allowed mass of a viable EOS exceed this mass,
\begin{equation} \label{mass_constraint}
M_{\rm max} \geq M^{\rm J0740}.
\end{equation}
In our models we ignore stellar rotation, but note that uniformly rotating NSs can support maximum masses up to about $20\%$ larger than their nonrotating counterparts (see, e.g., \cite{CooST94,CooST94b}).  

We show the maximum allowed masses for our sample of EOSs in Fig.~\ref{fig:M_max}, where (blue) dots indicate that the EOS passes the constraint within the smaller error limits, (green) triangles pass within the larger error limits, and (red) crosses represent EOSs that are ruled out, at least as long our other assumptions hold (e.g.~regarding rotation).

\subsection{Stellar radius}
\label{sec:radius}

\begin{figure*}
    \centering
    \includegraphics[width=0.45\linewidth]{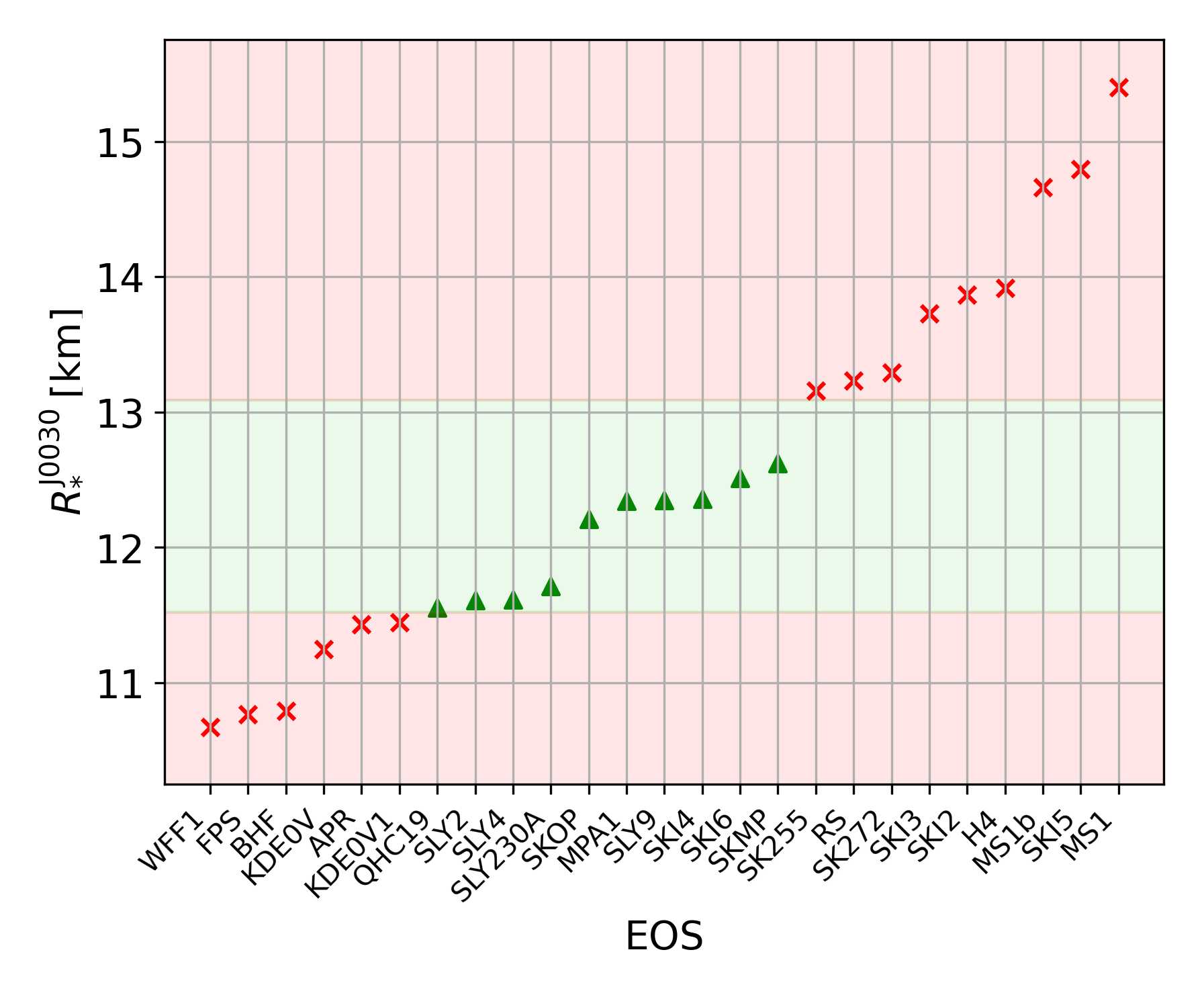}
    \includegraphics[width=0.45\linewidth]{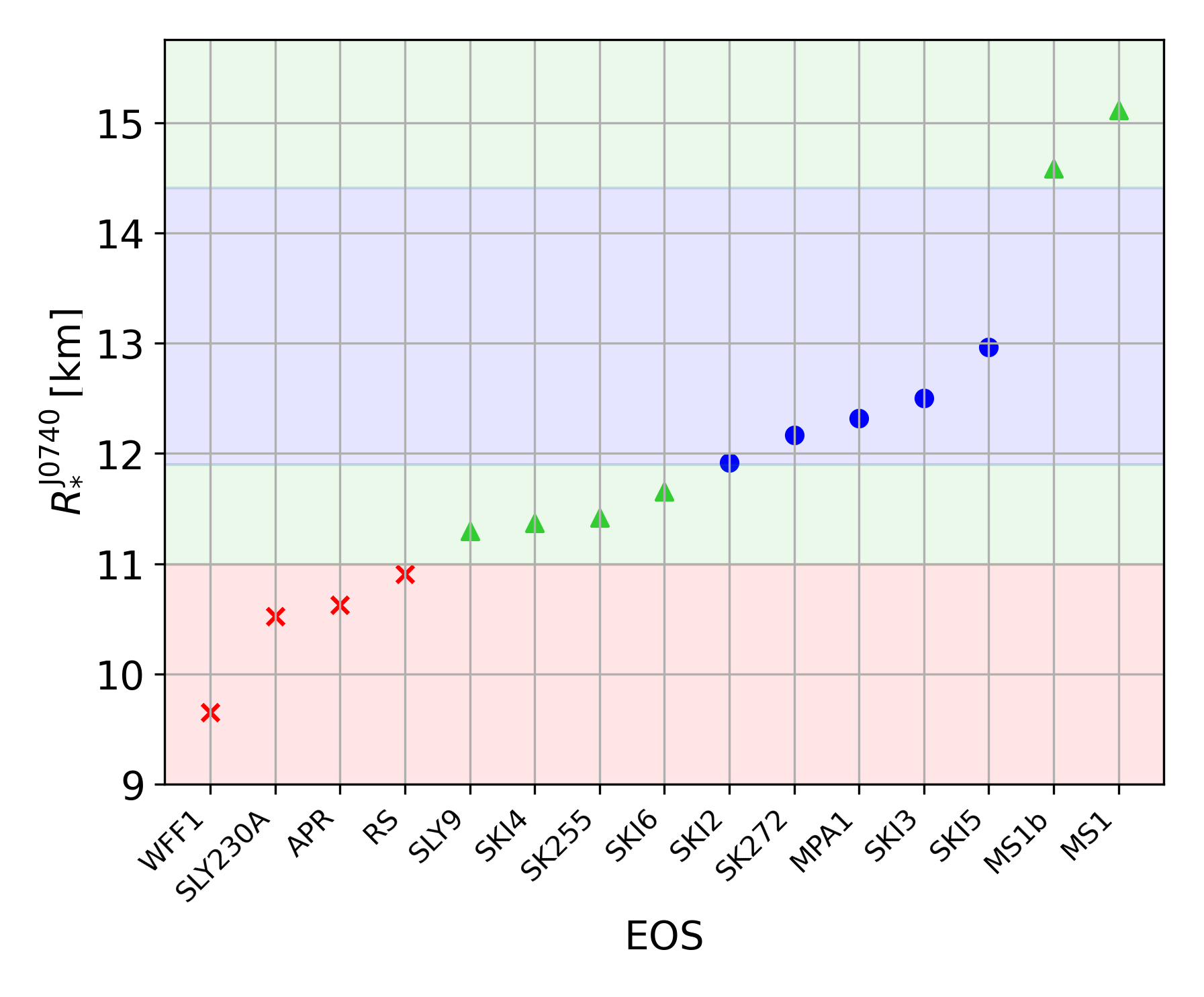}
    \caption{Numerical values of the stellar radius computed for a star of mass $M = 1.4 M_\odot$  representing J0030+0451 (left panel) and $M = 2.08 M_\odot$, representing J0740+6620 (right panel).  The right panel includes fewer EOSs, because several EOSs included in the left panel cannot support a star of mass $M = 2.08 M_\odot$ (see Section \ref{sec:m_max}).  The symbols (and colors) have the same meaning as in Fig.~\ref{fig:M_max} and indicate whether the numerical value of the radius agrees with the radii (\ref{R_J0030}) and (\ref{R_J0740}) observed by NICER to within the provided credibility limits. Only $95\%$ limits are provided in \cite{Raaetal21}, and thus the left panel omits symbols (and colors) for a higher credibility limit.   }
    \label{fig:R_constraints}
\end{figure*}

A second constraint on candidate EOSs results from comparisons between X-ray emission models and observations from NICER, and is realized as constraints on the radii of specific NSs. 

The radius of J0030+0451, which has a mass of $M^{\rm J0030} = 1.44^{+0.15}_{-0.14} M_\odot$,  is measured to be 
\begin{equation} \label{R_J0030}
R^{\rm J0030}_{*} = 12.33^{+0.76}_{-0.81} \mbox{km},
\end{equation}
where the error bars represent $95\%$ credibility limits
(see \cite{Raaetal21}). The radius of PSR J0740+6620, with the mass given by (\ref{M_J0407}) is
\begin{equation} \label{R_J0740}
R^{\rm J0740}_{*} = 12.92^{+5.65}_{-1.93} \mbox{km},
\end{equation}
again with $95\%$ credibility, or $R^{\rm J0740}_{*} = 12.92^{+1.49}_{-1.02} \mbox{km},$ to within $68\%$ limits (see \cite{Ditetal24}). 

We now construct stellar models for our sample of EOSs and for the above NS masses and compare the resulting numerical radii with the values quoted above.  For simplicity we adopt the stellar masses cited above in our analysis, and ignore the effects of the error in the stellar mass on the numerical values of the radius.  We show our results in Fig.~\ref{fig:R_constraints}, where we use the same symbols (and colors) as in Fig.~\ref{fig:M_max} to indicate whether or not a given EOS passes the constraint.

\subsection{Tidal deformability}
\label{sec:Lambda}

\begin{figure}  
        \includegraphics[width=0.48\textwidth]{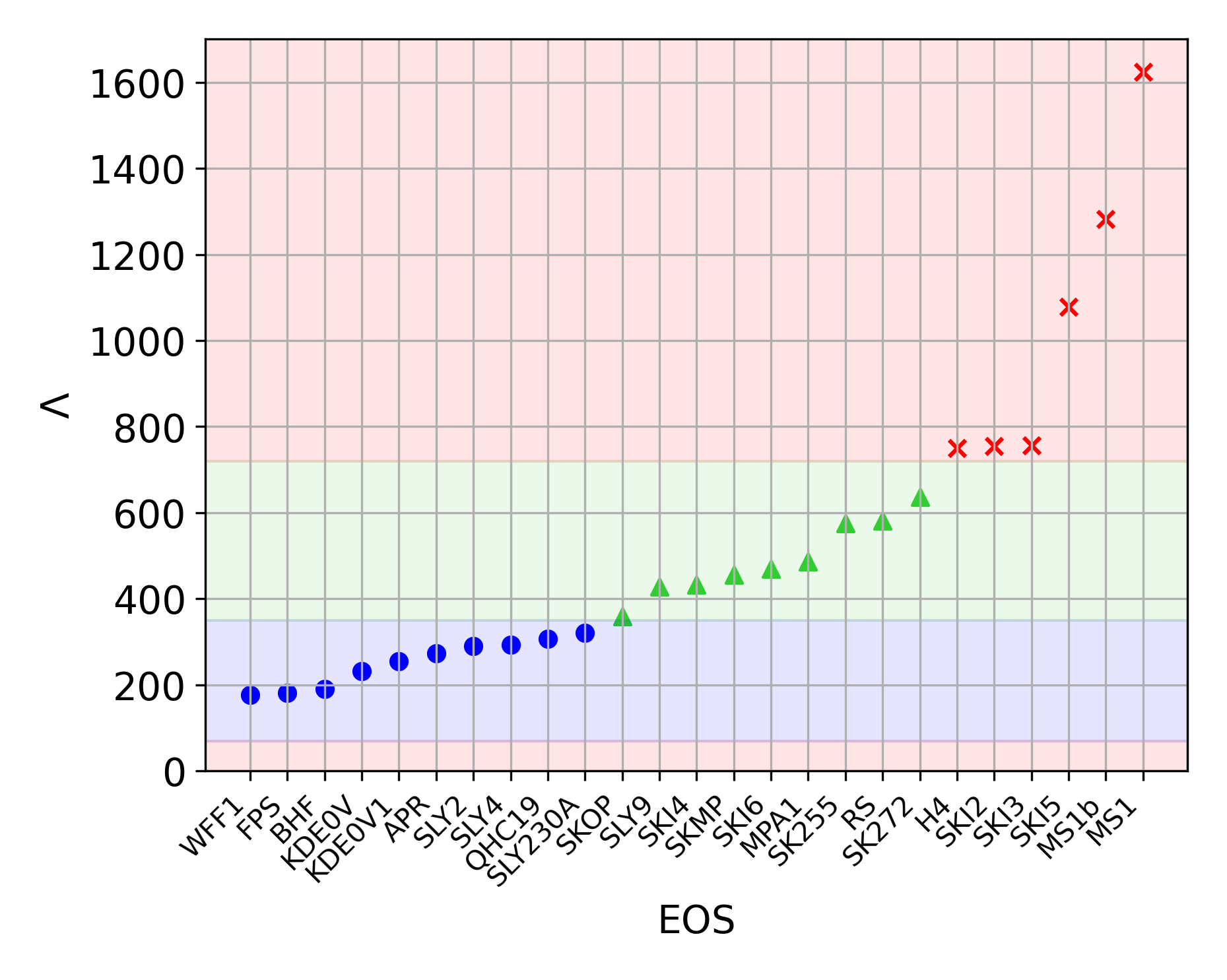}
        \caption{Values of the dimensionless tidal deformability parameter $\Lambda$ for a NS of mass $M = 1.4 M_\odot$ for our sample of EOSs.  As in Figs.~\ref{fig:M_max} and \ref{fig:R_constraints}, the symbols (and colors) indicate the level of agreement of our computed values with the value (\ref{Lambda}) observed in GW170817.}
        \label{fig:Lambda_constraints}
\end{figure}

A third constraint on candidate EOS results from the stellar tidal deformability.  

The tidal deformation of a NS in binary orbit affects the binary inspiral and hence the emitted GW signal.  To date, this has been observed only once, namely in the binary NS event GW170817, providing us with 
\begin{equation} \label{Lambda_GW170817}
\Lambda = 300^{+420}_{-230}
\end{equation}
(see \cite{Abbetal19}).  Here the error bars represent the $95\%$ confidence limits, while the $50\%$ upper bound on $\Lambda$ is given as 350. 

We calculated $\Lambda$ for our sample of EOSs by first computing the moment of inertia $I$ for stars of mass $M = 1.4 M_\odot$, consistent with the stellar masses in the binary GW170817, using the slow-rotation formalism of \cite{Har67}.  We then adopt the $I$-Love-$Q$ relationships of \cite{YagY13}, where the dimensionless tidal deformability parameter $\Lambda$ is defined as $\Lambda \equiv  \lambda  / M^{5}$ in terms of the tidal Love number $\lambda$. Specifically, we compute the dimensionless tidal deformability parameter $\Lambda$ from 
\begin{multline} \label{Lambda}
\ln{\Bar{I}} = 1.47+0.0817\ln{\Lambda}+0.0149(\ln{\Lambda})^{2}\\+(2.87 \times 10^{-4})(\ln{\Lambda})^{3}-(3.64 \times 10^{-5})(\ln{\Lambda})^{4}
\end{multline}
where $\Bar{I} \equiv I/M^{3}$ (see Eq.~(54) in \cite{YagY13} with coefficients provided in their Table I; also \cite{ArnB77,HinLLR10}).  

In Fig.~\ref{fig:Lambda_constraints} we compare our computed values of $\Lambda$ with the observed value (\ref{Lambda_GW170817}), again using the same symbols (and colors) as in Figs.~\ref{fig:M_max} and \ref{fig:R_constraints} to indicate the level of agreement. 

With improved sensitivity of existing GW observatories, and certainly with next-generation detectors, we are likely to observe significantly more binary-NS mergers, so that constraints on the tidal deformability will likely improve in the near future.

\section{New constraints from orbits of PBHs inside NSs}
\label{sec:PBH}

We now turn to hypothetical future observations of GWs emitted by PBHs orbiting inside NSs, and explore how such observations could supplement existing constraints on the nuclear EOSs and provide complementary information.

\subsection{Orbital precession and GW beats}
\label{sec:review}

If PBHs exists, and possibly make up a significant fraction of the Universe' dark matter content, some are likely to collide with stars.  If the PBH loses a sufficient amount of energy  in the initial collision, it remains gravitationally bound to the star and will ultimately be swallowed completely, possibly after many more transits.  As discussed by a number of different authors (see, e.g., \cite{CapPT13,AbrBW18,GenST20,AbrBUW22,BauS24b,CaiBK24}), this process is most likely to happen for NSs.  Once inside the NS, the motion of the PBH is governed by both dynamical and secular processes.  

Dynamically, the orbit results from the gravitational force -- and corresponding potential $\Phi$ -- exerted by the star on the PBH.  Note that, in the stellar interior, the gravitational potential is {\em not} equivalent to that of a point-mass, $\Phi \propto R^{-1}$, so that the orbits are {\em not} Keplerian.

In Newtonian gravity the gravitational potential of a constant-density star is that of a harmonic oscillator, with $\Phi \propto R^2$.  According to Bertrand's theorem (\cite{Ber73}; see also \cite{Gol80} for a textbook treatment), this is the {\em only} case other than the point mass potential $\Phi \propto R^{-1}$ that always leads to closed orbits.  The orbits can be described as ovals centered on the center of the star, and they do not precess, so that each polarization of the emitted GW signal will remain constant.  The orbital period $\tau_{\rm orb}$ is similar to the dynamical timescale of the host star, and therefore on the order of a millisecond for a NS.  Accordingly, the frequency of the GW signal, which, in analogy to radio waves, we will later refer to as the {\em carrier frequency}, is typically on the order of a few kilohertz.  Detecting such a frequency is a challenge for existing GW detectors, but may be possible with next-generation detectors like the Cosmic Explorer \cite{ce}, the Einstein Telescope \cite{et}, or special-purpose instruments like the Neutron Star Extreme Matter Explorer (NEMO, \cite{NEMO}). We also note that PBHs orbiting inside NSs are sources of {\em continuous} GW signals, meaning that they emit many GW cycles at almost constant frequency, as the orbital decay timescale is much longer than the dynamical timescale (see, e.g., \cite{BauS24b} as well as the discussion below).  This feature facilitates their detection (see, e.g., \cite{Ril23}). 

\begin{figure}
    \centering
    \includegraphics[width=0.95\linewidth]{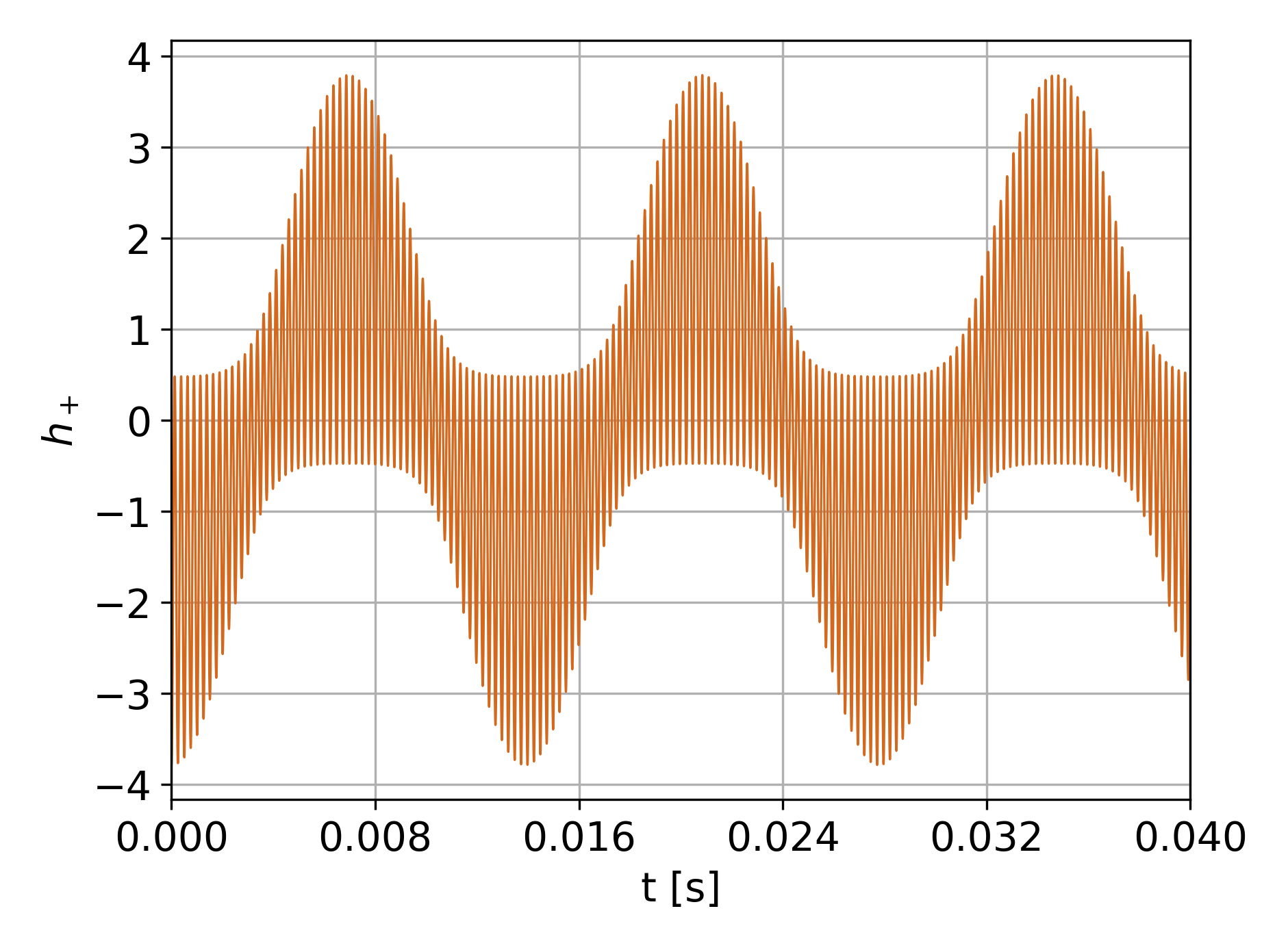}
    \caption{The ``+" polarization of a hypothetical GW signal emitted by a PBH inside a NS.  The short ``carrier" oscillations are caused by the PBH's orbit, with a period on the order of milliseconds, similar to the NS's dynamical timescale, while the modulation of these oscillations results in what we refer to as GW beats.  The size of the modulation depends on the eccentricity of the PBH's orbit.  The $y$-axis is in arbitrary units, as the amplitude of the wave depends on the PBH mass and the distance to the star.}
    \label{fig:GW}
\end{figure}

Both general relativistic effects and density inhomogeneity, however, cause the potential to depart from that of a harmonic oscillator, resulting in a precession of the orbit (see \cite{BauS24} for a detailed discussion).   This precession also affects the polarization of the emitted GW signal as follows.  At one moment in time the signal may be dominated by the $h_+$ polarization, say, but once the orbit has rotated by $45 \degree$ it will be dominated by the $h_\times$ polarization, only to be dominated by $h_+$ again once the orbit has turned by $90 \degree$.  Observing this signal in a fixed polarization would therefore result in a modulation of the GW amplitude that we refer to as {\em GW beats}.  It turns out that the rate of precession is quite sensitive to the stellar structure, and hence the EOS, so that an observation of these GW beats might provide stringent new constraints on the EOS.  In the following we will refer to the beat period, corresponding to a complete cycle in the GW signal, and hence an orbital pericenter advance by $180 \degree$, as $\tau_{\rm beat}$.  We will often express the beat period in terms of the orbital period $\tau_{\rm orb}$, yielding the number of carrier cycles per beat, 
\begin{equation}
{\mathcal N}_{\rm GW} \equiv \frac{2 \tau_{\rm beat}}{\tau_{\rm orb}}.
\end{equation}
Here the factor of two accounts for the fact that the GW signal completes two cycles per orbital revolution of the PBH.   We show an example of a hypothetical GW form in Fig.~\ref{fig:GW}.

In addition to the above dynamical processes, the PBH's orbit also evolves on secular timescales due to the retarding effects of dynamical friction, accretion drag, and radiation reaction forces.  As a result, the orbit slowly shrinks, and the PBH approaches the stellar center while accreting stellar material.  For much of this inspiral the orbit's eccentricity changes only very little (see Fig.~10 in \cite{BauS24b}), acting as an adiabatic invariant.   The rate of accretion is very small for small PBHs, but increases as the mass of the PBH increases, and ultimately induces stellar collapse in which the PBH swallows the entire star in ``one last gulp".  In \cite{BauS24b} we found that the timescale for this growth is similar to, but somewhat smaller than what we had previously estimated to be a nearly universal maximum survival time of a NS harboring a PBH with initial mass $m_{\rm  PBH}(0)$, 
\begin{equation} \label{tau_max}
\tau_{\rm max} \simeq 10^6 \, \mbox{s} \, \left( \frac{10^{-10} M_\odot}{m_{\rm PBH}(0)} \right)
\end{equation}
(see Eq.~(17) in \cite{BauS21} as well as Section VI.C in \cite{BauS24b}, and compare similar estimates in \cite{GenST20,BauS24b}).  For sufficiently small PBHs the secular inspiral timescale $\tau_{\rm sec}$, comparable to (\ref{tau_max}), is orders of magnitude larger than the dynamical processes discussed above, $\tau_{\rm sec} \gg \tau_{\rm beat} > \tau_{\rm orb}$.  We may therefore ignore these retarding forces in our simulated orbits, as they have negligible effect on the timescale that we are interested in, nor do they affect the nature of the signal as a continuous GW source. 

\subsection{Observational scenarios}
\label{sec:scenarios}

We now discuss some possible scenarios to explore how the observation of GW beats could be used to constrain the nuclear EOSs.  We show an example of such a hypothetical GW in Fig.~\ref{fig:GW}.  Some conclusions could be drawn almost immediately, while others require more analysis.

We first observe that the PBH's orbital frequency $\omega_{\rm orb}$ can be identified directly from the ``carrier" frequency.  In our specific example we chose $\omega_{\rm orb} = 1.3 \times 10^4$ rad s$^{-1}$, which we will adopt as our fiducial observed orbital frequency $\omega_{\rm fid}$ for our discussion here.  Furthermore, the size of the modulations is directly related to the eccentricity of the orbit.  In Fig.~\ref{fig:GW} we chose a rather large eccentricity for purposes of illustration, with $f_\varphi = 0.25$, since this results in clearly visible beat patterns.  For simplicity, we will instead assume a nearly circular orbit in the following discussion, since the orbital frequency $\omega_{\rm orb}$ is then very similar to the circular orbital frequency $\omega_{\rm cir}$, and can hence be determined directly from the stellar model without the need of integrating eccentric orbits numerically, as sketched in Section \ref{sec:OV}.  Otherwise, however, our arguments can be applied similarly to eccentric orbits. In Fig.~\ref{fig:GPP_omegavsR_1.4} we show $\omega_{\rm cir}$ as a function of the orbital radius $R$ for different EOSs and two different stellar masses, and include the fiducial observed value $\omega_{\rm fid}$ as the orange horizontal line.

Even a single observation of $\omega_{\rm orb}$ by itself may provide some immediate constraints.  Specifically we note that some EOSs, e.g.~MS1, would be unable to support $\omega_{\rm fid}$ for neither one of the two masses shown in Fig.~\ref{fig:GPP_omegavsR_1.4}, nor any mass in between, and could therefore be ruled out.  Other EOSs, however, could support circular orbits with $\omega_{\rm fid}$ at different orbital radii $R$ and different masses $M$, as demonstrated by the intersections between the curves $\omega_{\rm cir}(R)$ and $\omega_{\rm fid}$ in Fig.~\ref{fig:GPP_omegavsR_1.4} (marked by the dots).\footnote{Many EOSs allow {\em two} different radii $R$ for a given orbital frequency $\omega_{\rm orb}$, see Fig.~\ref{fig:GPP_omegavsR_1.4}.  Observationally, these two radii should be distinguishable, because one, for the larger radius, occurs before $\omega_{\rm orb}$ takes on its maximum value $\omega_{\rm max}$ during the PBH's inspiral toward the stellar center, while the other, for smaller $R$, occurs after.  For the purpose of our discussion here we always pick the smaller value of $R$.}   Constraining the EOS therefore amounts to simultaneously determining the orbital radius $R$, the stellar mass $M$, and the EOS -- which we may think of as a three-dimensional parameter space.  (For general orbits, the eccentricity would constitute a fourth dimension, but it can be constrained from the size of the beat modulations, as discussed above, or by the shape of the carrier waveform, as in compact binaries.  It is then clear that an observation of $\omega_{\rm orb}$ and $\NGW$ alone is not sufficient to determine all three of these parameters, and that some additional information is required.

\subsubsection{Known stellar mass}
\label{sec:knownmass}

\begin{figure}[t]  
        \includegraphics[width=0.48\textwidth]{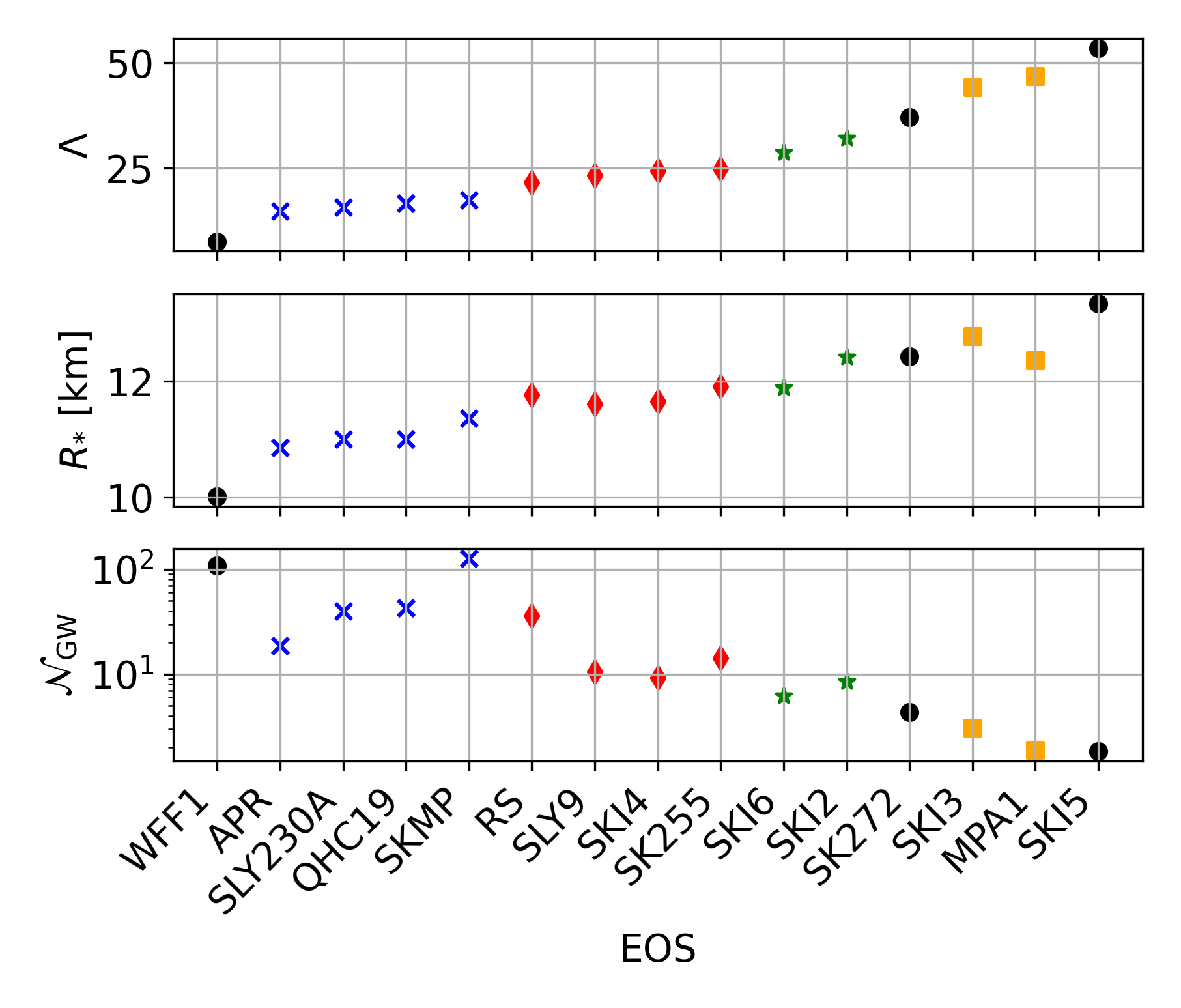}
        \caption{Values for the tidal deformability $\Lambda$, the stellar radius $R_*$, and $\NGW$ for an orbit with $\omega_{\rm fid} = 1.3 \times 10^4$ s$^{-1}$ in a star with mass $2.0 M_\odot$.  The four highlighted groups -- (blue) crosses, (red) diamonds, (green) stars, and (yellow) squares -- have quite similar values of $\Lambda$ and $R_*$, but some of them differ significantly in $\NGW$.  For example, APR and SLY230A would be difficult to distinguish from $\Lambda$ and $R_*$ alone, which are very similar for a $M = 2.0 M_\odot$ star governed by these two EOSs, while $\NGW$ differs by about a factor of two.}
        \label{fig:fixed_mass}
\end{figure}

Mostly for illustrative purposes, we first consider the unlikely scenario that the mass of the host star is known, say $M = 2.0 M_\odot$.  In this case, the observed orbital frequency $\omega_{\rm orb}$ corresponds to a specific orbital radius for a given EOS; for our adopted fiducial value $\omega_{\rm fid}$ these radii are marked by the dots in the right panel of Fig.~\ref{fig:GPP_omegavsR_1.4}.  For these radii we can then integrate the equations of motion for the PBH for several orbits, as discussed in Section \ref{sec:orbits}, from which we can determine $\NGW$.  

In Fig.~\ref{fig:fixed_mass} we plot our results for $\NGW$ for the different EOSs, together with the corresponding values of the tidal deformability $\Lambda$ and stellar radius $R_*$.  As can be seen from the figure, some EOSs would be difficult to distinguish from their values of $\Lambda$ and $R_*$ alone, because they take very similar values for a $M = 2.0 M_\odot$ star, while, by contrast, some of the values of $\NGW$ differ significantly.  While knowing the mass of the NS independently is quite unlikely, this example already demonstrates how an observation of GW beats from PBHs might provide complementary information on the nuclear EOS that could break degeneracies between existing constraints.

\subsubsection{Unknown stellar mass}
\label{sec:unknownmass}

\begin{figure}
    \centering
    \includegraphics[width=0.99\linewidth]{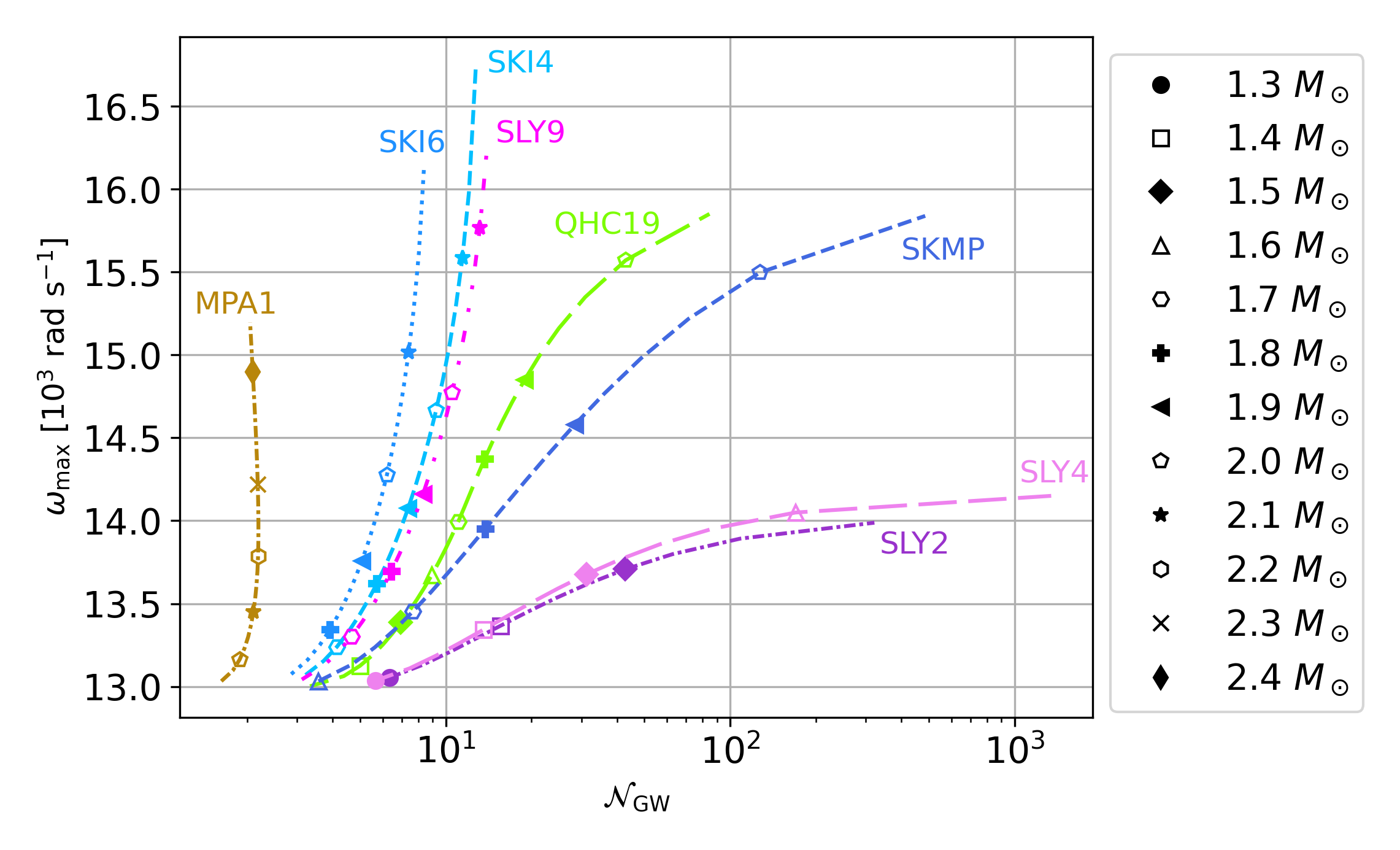}
    \caption{Sequences of $\omega_{\rm max}$ versus $\NGW$ for those EOSs that have not been ruled out by the existing constraints discussed in Section \ref{sec:constraints}.  For each EOS we consider masses between $M_{\rm min} = 1.3 M_\odot$ and the maximum mass allowed by that specific EOS (see Table \ref{tab:EOS_summary}).  For each mass (and each EOS), we then compute the maximum orbital frequency $\omega_{\rm max}$ as well as $\NGW$ for the observed value $\omega_{\rm obs}$ (in this case our fiducial value of $\omega_{\rm fid} = 1.3 \times 10^{4}$ Hz).  The resulting curves in the $\omega_{\rm max}$ versus $\NGW$ diagram are then parametrized by the stellar mass, as indicated by the symbols along each line.  With very few exceptions these lines do not intersect, so that an observation of $\NGW$ together with $\omega_{\rm max}$ could uniquely determine both the EOS and the stellar mass.}
    \label{fig:free_mass}
\end{figure}

We now turn to the much more likely scenario that no independent information about the host star is available, and that its mass in particular is unknown.  As we discussed before, an observation of $\omega_{\rm orb}$ and $\NGW$ alone is then insufficient to determine the orbital radius $R$, the stellar mass $M$, and the EOS.  However, $\omega_{\rm orb}$ and $\NGW$ correspond to only one ``snapshot" of the PBHs inspiral towards the stellar center, i.e.~for one orbital radius, and we may complement this information with some other information about the entire inspiral signal.

A number of different characteristics could be adopted in this approach, for example multiple values of $\omega_{\rm orb}$ and their corresponding values of $\NGW$.  Instead, we here explore using a rather simple characteristic based on the maximum circular orbital frequency $\omega_{\rm max}$.  We assume that at least a significant fraction of the PBH inspiral towards the stellar center has been observed.\footnote{While the inspiral secular timescale (comparable to Eq.~(\ref{tau_max})) greatly exceeds the typical orbital timescale ($\sim$ ms), it is sufficiently short that the PBH can be tracked as it inspirals towards the center.}   During this inspiral the orbital radius shrinks, so that the orbital frequency $\omega_{\rm orb}$ follows one of the lines shown 
in Fig.~\ref{fig:GPP_omegavsR_1.4}.  In particular, $\omega_{\rm orb}$ should take a maximum value $\omega_{\rm max}$ for the stars considered here, which we will adopt in the following.

An example of how $\omega_{\rm max}$ together with $\NGW$ could be used to determine both the EOS and the stellar mass is shown in Fig.~\ref{fig:free_mass}.  Specifically, we consider the eight different EOSs that have not yet been ruled out by the existing constraints discussed in Section \ref{sec:constraints}.  For each of these EOSs we consider stellar masses ranging from $M_{\rm min} = 1.3 M_\odot$ to the maximum allowed mass of the respective EOS (see Table \ref{tab:EOS_summary}), and compute, for these masses, the value of $\omega_{\rm max}$, as well as $\NGW$, adopting our fiducial value $\omega_{\rm fid}$ as an example of an observed orbital frequency.   For each EOS, this results in a line in the $\omega_{\rm max}$ versus $\NGW$ diagram, parametrized by the stellar mass.  The latter is indicated by the different symbols in Fig.~\ref{fig:free_mass}.  

We first observe that two pairs of lines, for SKI4 and SLY9 as well as SLY2 and SLY4, lie close together and would be difficult to distinguish.  This is hardly surprising, however, because the EOS themselves are very similar, as we discussed at the end of Section \ref{sec:EOS}.  Other than these two pairs, there is only one intersection between the remaining lines.  Threrefore, an observation of both $\NGW$ and $\omega_{\rm max}$, corresponding to a single point in Fig.~\ref{fig:free_mass},  would in all likelihood uniquely determine both the EOS and the stellar mass.

In the above discussion we assumed, for simplicity, that the PBH orbit inside the NS is nearly circular.  Instead, it is probably more realistic to assume the orbit to have an appreciable eccentricity (see \cite{BauS24b}), which will also lead to larger modulations of the GW polarizations.  As we discussed above, the eccentricity can then be determined from the size of the GW modulation and the shape of the carrier waveform, and it is likely that the eccentricity will change only little during the PBH's inspiral to the stellar center (see Fig.~10 in \cite{BauS24b}).  For such eccentric orbits, constraints on the NS EOS can be established exactly as in our discussion above, with the only difference that the orbital frequency $\omega_{\rm orb}(R)$ can no longer be determined directly from the background NS model alone (as discussed in Section \ref{sec:OV}), but instead has to be found by integrating the relativistic geodesic equations for a few orbits.  This adds one more step in the analysis, but does not affect the overall strategy.

Finally, we reiterate that the above procedure is based on only one characteristic of the inspiral signal, namely $\omega_{\rm max}$, and that it could be either replaced or complemented with other characteristics, e.g.~values of $\omega_{\rm orb}$ and $\NGW$ at different orbital radii.

\section{Summary and Discussion}
\label{sec:discussion}

In summary, we explore how GWs emitted by PBHs orbiting inside NSs may provide constraints on the nuclear EOS together with information about the host NS.  Orbits of PBHs inside NSs precess at a rate that strongly depends on the NS structure, and hence on the nuclear EOS.  As the orbit precesses, the maximum amplitude of the GW oscillates between the two polarizations, leading to beat phenomena when observing one of the polarizations.  A future observation of these beat frequencies can therefore provide valuable information about the nuclear EOS, potentially complementing existing constraints.  In \cite{BauS24}, we discussed this scenario in the context of simple polytropic EOSs, while here we consider realistic nuclear EOSs.

For each EOS in our sample we construct models of NSs for different masses $M$ and compute orbits of PBHs inside these stars.  In particular, we track the orbital precession frequency by determining the rate of pericenter advance for each orbit, and then compute the associated GW beat frequency.  We then discuss how a hypothetical future GW detection could be used to distinguish between different EOS, stellar masses, and orbital radii.  We observe that an ``instantaneous" observation of GW emitted at a single orbital radius is not sufficient to determine all three of the above.  Supplementing such an observation with some additional information about the entire inspiral, however, e.g.~the maximum orbital frequency $\omega_{\rm max}$, should be sufficient to simultaneously determine the EOS, the stellar mass, as well as the orbital radius.  While we focus on nearly circular orbits in our discussion here for simplicity, a similar analysis can be performed for orbits with larger eccentricity.  Such an analysis will be facilitated by the fact that the eccentricity is expected to remain nearly constant during the inspiral (see \cite{BauS24b}).

The event rates for the capture of PBHs by NSs are expected to be small, which makes a detection with current GW detectors unlikely.  A future detection with improved detectors, however, could complement existing constraints on the EOS, and could provide highly valuable information on the nature of nuclear matter as well as neutron stars.


\acknowledgements

A.C.~and J.D.D.~gratefully acknowledge support through Undergraduate Research Fellowships at Bowdoin College.  This work was supported in part by National Science Foundation (NSF) grants PHY-2010394 and PHY-2341984 to Bowdoin College, as well as NSF grants PHY-2006066 and PHY-2308242 to the University of Illinois at Urbana-Champaign.


%

\end{document}